\documentclass{emulateapj}
\usepackage{epsfig}
\usepackage{lscape}
\newcommand{\ch}{{\it Chandra}}
\newcommand{\xmm}{{\it XMM}}
\newcommand{\xmmn}{{\it XMM-Newton}}
\slugcomment{Sep. 10, 2005}
\begin{document}

\title{Type II Quasars from the SDSS: IV. \ch\ and \xmmn\ Observations Reveal Heavily Absorbed Sources.}

\author{
Andrew Ptak\altaffilmark{1}, 
Nadia L. Zakamska\altaffilmark{2}, 
Michael A. Strauss\altaffilmark{2}, \\
Julian H. Krolik\altaffilmark{1}, 
Timothy M. Heckman\altaffilmark{1},  
Donald P. Schneider\altaffilmark{3}, Jon Brinkmann\altaffilmark{4} 
}
\altaffiltext{1}{Department of Physics and Astronomy, Johns Hopkins
  University, 3400 North Charles Street, Baltimore, MD 21218-2686} 
\altaffiltext{2}{Princeton University Observatory, Princeton, NJ 08544}
\altaffiltext{3}{Department of Astronomy and Astrophysics, 504 Davey
  Laboratory, Pennsylvania State University, University Park, PA
  16802} 
\altaffiltext{4}{Apache Point Observatory, P.O. Box 59, Sunspot, NM 88349} 

\begin{abstract}
We are carrying out sensitive X-ray observations with \ch\ and \xmm\
of type II quasars selected from the Sloan Digital Sky Survey based on
their optical emission line properties.  In this paper we present the
first results of our program.  We present observations of four objects
at redshifts $0.4<z<0.8$ and an analysis of the archival data for four
additional objects in the same redshift range.  Six of the eight were detected
in X-rays; five of them have sufficient signal to derive spectral
information.  All of the detected sources have 
intrinsic luminosities 
$L_{2-10 \rm \ keV} > 5 \times 10^{43}$ erg s$^{-1}$; the five
with sufficient counts for spectral fitting show evidence
for significant absorption ($N_H\ga$ a 
few$\times 10^{22}$ cm$^{-2}$).  At least three of the objects
likely have $N_H>10^{23}$ cm$^{-2}$; some may be Compton-thick
($N_H > 10^{24} \rm \ cm^{-2}$).  
In the five objects for which we could fit spectra, the slopes tend to
be significantly flatter than is typically observed in AGN; it is
possible that this is due either to reprocessing of the nuclear emission or 
to a line of sight that passes through patchy absorption.
\end{abstract}


\section{Introduction}

Type II Active Galactic Nuclei (AGN) are defined as those AGN
lacking broad optical emission lines.  In many cases, particularly
at low luminosity (i.e., Seyfert galaxies), the
absence of broad optical emission lines has been shown to be due
to obscuration, i.e., the central regions are blocked from the observer
by a high column density of gas and dust.   In the standard
unification model, the absorbing material is toroidal, resulting in highly
anisotropic optical, UV and X-ray emission, while the strong narrow
emission lines are produced above and below the obscuring torus in
matter illuminated by the central engine.  The detection of broad
emission lines in polarized, and hence scattered, light shows that at
least in some cases the lack of broad lines is due to orientation
effects \citep{anto93}.   Although low-luminosity obscured AGN
(type II Seyfert galaxies) are abundant 
in the local universe \citep{kauf04, hao05}, until recently only a
few type II quasar candidates were known (e.g., \citealt{klei88}).

A key question is therefore to what extent the standard model applies
to type II quasars.  This is the fourth in a series of papers using
a sample selected on the basis of Sloan Digital Sky Survey (SDSS)
data to study the numbers and properties of high-luminosity type II
AGN, type II quasars.
Our method is based on the supposition (as posited in the standard
model) that most of the narrow optical line emission originates outside the
obscuring torus, and is therefore expected to be roughly isotropic.
We have selected several hundred type II quasar 
candidates from the spectroscopic database of the SDSS based on their
narrow emission line properties 
(\citealt{zaka03}, hereafter Paper I).  We are now conducting sensitive
multi-wavelength observations of these objects to determine the
range of their properties and to understand how they compare to type II
AGN in general. In this paper we present the first results of our
\ch\ \citep{weiss02} and \xmm\ \citep{jansen01} programs.  Our main goal is the determination of
the X-ray spectral properties of this sample: their X-ray
luminosities, continuum spectral slopes, and line-of-sight obscuration.

Other methods have been used recently to search for type II AGN.
Deep hard X-ray (i.e., E $>$ 2 keV) surveys efficiently detect AGN
candidates, including potentially obscured ones.  
For example, the X-ray
background in the 1--6~keV band has been largely resolved into
AGN at redshifts around
$z\sim 1$ \citep{horn01}.  However, while many of these X-ray
sources show only narrow emission lines in their optical spectra, a
significant number of X-ray sources show broad emission lines or no
emission lines whatsoever (e.g., \citealt{barg03, szok04, barg05,
heck05}, and \citealt{matt02} for an earlier review).  In addition,
many X-ray selected 
AGN candidates are very faint in the optical (i.e., $R \ga 24$).
Consequently, it is difficult even to obtain complete redshift
information for samples selected in this manner, much less a
complete classification of their optical spectroscopic types.

Alternatively Type-2 AGN are often detected in large-area hard
X-ray surveys \citep{della04, cacc04, revn04} and X-ray
follow-ups to very hard (E $>$ 10 keV) X-ray all-sky surveys
\citep{sazo05}.  These methods 
are promising since the detected objects tend to be (or are selected
to be) bright in X-rays and often also in the optical, allowing for
spectroscopic classification and redshift determination.  However,
while several Type-2 quasars have been found (i.e., with $L_X > 10^{44} \rm
\ ergs \ s^{-1}$), these 
sources tend to be mostly Seyfert 2 galaxies at low redshifts
(particularly in the very hard X-ray surveys since the high background of
non-imaging detectors results in high flux limits).  \citet{seve05}
discuss a related approach in which a field with known Extremely Red
Objects (EROs) was surveyed with \xmmn, and several EROs were found
to harbor obscurred AGN and quasars.
Likewise, a considerable effort is being applied to use mid-IR colors
to select AGN \citep{haas04,lacy04,ster04}, but the completeness and
efficiency of the selection of type II AGN by this method has not yet been
established.   
Along these lines, \citet{mart05} selected type II quasars at $z
\sim 2$ using the Spitzer First Look Survey data along with radio
fluxes to find type II AGN and exclude starburst galaxies.  This study
suggests that the type II AGN fraction is $\sim 50\%$ at $z \sim 2$,
but the uncertainties are large due to small number statistics and
model-dependent assumptions.

In Section \ref{sec_sam} we review in detail the properties of the
SDSS type II AGN sample. Section \ref{sec_obs} describes our observations
and data reduction, Section \ref{sec_spec} describes the X-ray spectral 
analysis,
Section \ref{sec_individ} gives details on individual objects. We
discuss our results in Section \ref{sec_disc}, followed by a brief
summary in Section 7. An $h=0.7$,
$\Omega_m=0.3$, $\Omega_{\Lambda}=0.7$ cosmology is assumed
throughout. We frequently refer to the [OIII]$\lambda$5007\AA\ optical
emission line as simply [OIII]. Objects are identified
by their J2000 coordinates in Table \ref{t:params} (e.g., SDSS
J084234.94+362503.1) and shortened to {\it hhmm+ddmm} notation
elsewhere (SDSS~J0842+3625).

\section{Sample description}
\label{sec_sam}

The SDSS \citep{york00, stou02, abaz03,abaz04, abaz05} is an ongoing
optical survey to image about 10,000 deg$^2$ and obtain spectra of
about 10$^6$ galaxies and 10$^5$ quasars. Using the spectroscopic
database of the SDSS as of July 2002 (about $4\times 10^5$ spectra),
we identified 291 objects in the redshift range $0.3<z<0.8$ having
strong narrow emission lines with high-ionization line ratios
characteristic of an underlying AGN continuum (Paper I). Neither broad
emission lines nor the strong ionizing UV continuum are seen in these
spectra, suggesting that the central engine is blocked from the
observer.  
All these objects are therefore type
II AGN candidates based on their optical properties.  

Since the narrow lines are presumably
illuminated by the central engine similarly in type I and type II AGN,
narrow line luminosities can serve as a proxy for the nuclear 
luminosity. We used the [OIII]$\lambda$5007\AA\ emission line, which
is present in all of the SDSS spectra of our objects, and one of the
strongest optical emission lines in type II AGN \citep{oster89}. About
50\% of the  
objects in our sample (130/291) have [OIII]$\lambda$5007\AA\
luminosities in excess of 
$3\times 10^8 L_{\odot}$, similar to those of luminous unobscured
quasars ($-23>M_B>-27$). Based on this somewhat arbitrary criterion
we classified objects with $\log (L$[OIII]$/L_{\odot})>8.5$ as type
II quasars; those with lower luminosities we designated
type II Seyfert galaxies (Paper I). 

Using data from the IRAS all-sky survey we found that the mid-IR
luminosities of the objects in our sample reach a few$\times 10^{46}$
erg s$^{-1}$ \citep{zaka04}, and therefore
we directly confirmed the luminous nature of these objects. 
We have conducted spectropolarimetry of a subsample of the most
luminous objects \citep{zaka05} and have shown
that they harbor luminous blue broad-line quasars in their
centers. These results are consistent with the basic
orientation-based unification model of toroidal obscuration and
off-plane scattering \citep{anto93}, implying that the model can be
extended to include at least some high-luminosity AGN. 

We used the ROSAT All-Sky Survey (RASS; see Voges et al. 1999 and
references therein) data to investigate the X-ray
properties of the objects in our sample \citep{zaka04}. We found that type
II AGN are about 10 times less likely to be soft X-ray sources than
are type I AGN with the same redshift and [OIII] luminosity.
Furthermore, an unexpectedly high fraction (50\%) of the  
few objects with counterparts in RASS are radio-loud, in which
cases the soft X-ray emission may be associated with jet activity
rather than with the central engine \citep{urry04}. These findings indicate
either that type II AGN are significantly underluminous in X-rays compared
to type I AGN, or, more likely, that the soft X-ray emission is absorbed by
intervening material in type II AGN, consistent with the standard
unification model.  Similar conclusions were obtained by
\citet{vign04}, who investigated pointed ROSAT observations of the
fields of 16 objects from our sample and found that the majority of
objects are undetected in soft X-rays even in exposures much deeper
than those of RASS. Our program of sensitive X-ray observations
reaching well into the 2$-$10 keV band is intended to provide better
understanding of the low ROSAT detection rate of type II AGN, and to
measure their absorbing
column densities and the X-ray luminosities independently.

\section{Observations and Data Reduction}
\label{sec_obs}

Of the eight objects presented in this paper, two were observed as part
of our \ch\ proposal (ID 05701043), and two were observed as part of
our \xmm\ proposal (ID 020434). In addition, we checked all archival
pointings of \ch\ and \xmm\ within 15\arcmin\ of SDSS type II
AGN\footnote{using the High-Energy Astrophysics
Science Archive (HEASARC), http://heasarc.gsfc.nasa.gov} and found
that an additional four objects from Paper I were observed
serendipitously. The list of the objects, observation IDs, exposure
times and dates of the observations are given in Table
\ref{t:params}. Of the 
four objects observed serendipitously, two (SDSS~J0115+0015 and
SDSS~J0243+0006) do not meet our quasar luminosity criterion
($L$[OIII]$>3\times 10^8L_{\odot}$), so we classify them as Seyfert II
galaxies.  All eight objects in this paper are radio-quiet, as
determined by the ratio of their radio and [OIII] luminosities
(Zakamska et al. 2004).

We used XAssist \citep{ptak03} for initial processing of all
data. XAssist works similarly for both \ch\ and \xmmn\ data, with most
operations performed by the software packages CIAO (version 3.2.1
and CALDB 3.0.3) and XMMSAS (version 6.1.0). For each field, the data
were reprocessed to take advantage of the latest calibrations, sources
were detected, and the exposure was 
trimmed to remove background flares.  The size of each source on the
detector was estimated in order to determine appropriate source
extraction regions, typically
$\sim 2\arcsec$ regions (\ch) or $\sim 16-18\arcsec$ regions (\xmmn)
for on-axis point sources. For off-axis observations we used larger
source extraction regions, as described below for each individual
case. Spectral responses were calculated for the source spectra, and
background spectra were extracted from annuli centered on the sources,
with interloping sources excluded from the background regions. 

\section{Results of Spectral Analysis}
\label{sec_spec}

Of the eight objects in our sample, six were
detected in X-rays, and all are consistent with being point
sources.  We defined ``detection" as having enough counts
in excess of the expected background that the Poisson probability
of a spurious source at that location was $<  1\times 10^{-6}$
(corresponding to a $\sim 5\sigma$ criterion in Gaussian
statistics).  Since we are investigating specific locations rather
than sampling a large field, this limit is very conservative.
However, as discussed below, in only one case (SDSS~J0842+3625) would
a lower (yet reasonable) detection limit result in a detection, and in
that case the source detection is complicated by the position of the
source in the field.  For the purpose of estimating error bounds for the
luminosity, we treat all three weak sources, whether detected or not,
in the same fashion.

Sufficient counts for spectral fitting were detected from five objects
(with more than 300 counts; see Table 2).
We modeled the spectra with a
power-law continuum absorbed by the Galactic column density and an
absorber at the redshift of the source, resulting in three free
parameters (the column density $N_H$, X-ray photon index $\Gamma$:
$dN/dE \propto E^{-\Gamma}$, and the power-law
normalization).  The relation between HI column density and
opacity assumed solar abundances. 
Galactic absorption was derived from the HI map
\citep{dick90} using the {\sc nh} tool provided by the HEASARC.  The
Galactic neutral hydrogen column density does not exceed $4.8\times
10^{20}$ cm$^{-2}$ for any of the objects in our sample.

A simple absorbed power-law model is generally
inaccurate for very high column densities of the absorber ($N_H\ga
10^{23}$ cm$^{-2}$) because electron scattering significantly modifies
the observed spectrum.  We therefore used the approximate model by
\citet{yaqo97} that allows for scattering (``plcabs'' in {\sc xspec}).
When the Compton optical depth is significant, the flux that
emerges in any particular direction is strongly dependent on
the scatterer's geometry \citep{krol94, ghis94}; the plcabs model
assumes spherical symmetry, which 
is not necessarily appropriate (we expect that the
obscuration is toroidal).  Our hope is that this model provides
a slightly better estimate, particularly for Compton depths that
are order unity or less, than ignoring scattering altogether.
For low optical depths, this model is equivalent to simple
absorbed power-law models. 

The XMM-Newton spectra were binned in energy to 20 counts per bin to
allow the use of the $\chi^2$ statistic. For the two \ch\ 
observations, the expected background contribution to the source
region is negligible ($< 2$ counts in both cases), and we used the C
statistic \citep{cash79} for spectral fits, without binning and
subtracting the background.  
The parameter errors were computed at 90\% significance for one
interesting parameter
($\Delta \chi^2 \rm \ or \ C = 2.7$).  In the case of SDSS~J0801+4412,
which had only 40 detected photons,
we fixed the power-law photon index at $\Gamma=1.7$ (the typical
value found in X-ray surveys for AGN fit 
with an absorbed power-law model; e.g., \citealt{nand05,tozz05}).
Adopting values of $\Gamma = 1.4$  (the photon index of the X-ray
background in the 0.5-10.0 keV bandpass), or a steeper slope such as $\Gamma = 1.9-2.0$, resulted in similar fit parameters.

\begin{figure}
\epsscale{0.8}
\plotone{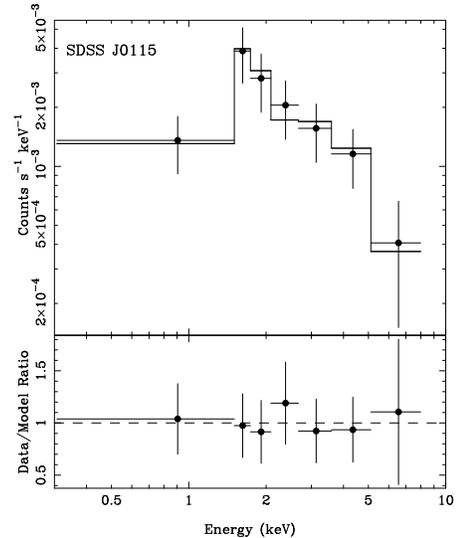}
\figcaption{The \ch\ ACIS spectrum of the serendipitously observed object
  SDSS~J0115+0015, fit with an absorbed power-law with the absorber at
  the redshift of the source and with observed energies plotted.  The
  residuals of the fit are shown in the bottom panel.\label{pic_0115}} 
\end{figure}

\begin{figure}
\epsscale{0.8}
\plotone{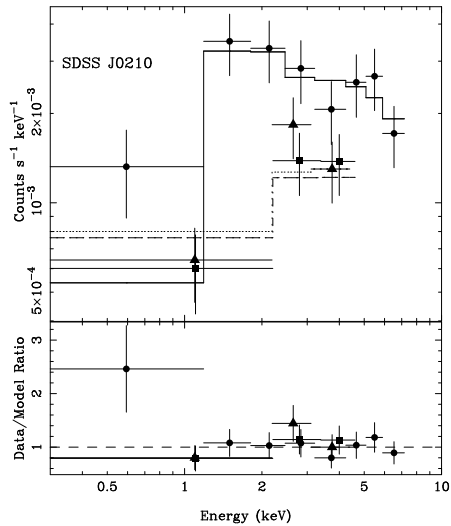}
\figcaption{The \xmmn\ spectra for SDSS~J0210$-$1001 fit with an
  absorbed power-law model. PN data are marked
  with circles, MOS1 data are marked with triangles, and the MOS2 are
  marked with squares. The solid, dotted, and dashed lines in the
  upper-panel show the best-fitting model for the PN, MOS1 and MOS2 data. \label{pic_0210}} 
\end{figure}

\begin{figure}
\epsscale{0.8}
\plotone{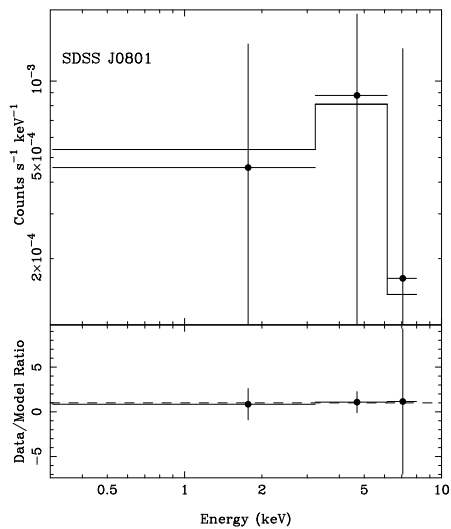}
\figcaption{The \ch\ spectra for SDSS~J0801+4412 fit with an absorbed
  power-law model. The spectrum was
  binned here for display purposes only.\label{pic_0801}}
\end{figure}

\begin{figure}
\epsscale{0.8}
\plotone{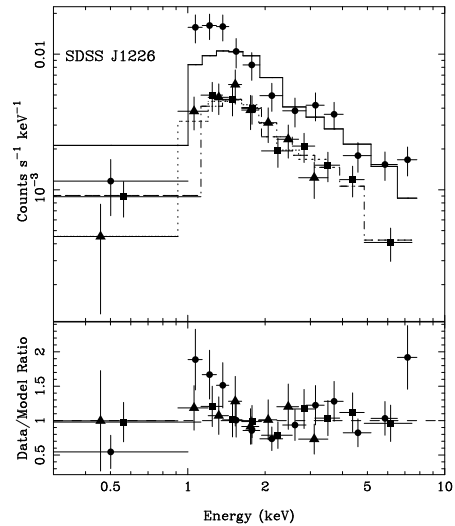}
\figcaption{\xmmn\ spectra of SDSS~J1226+0131 fit by an absorbed
  power-law model. Marks and lines are the same as in Figure
  \ref{pic_0210}. \label{pic_1226}} 
\end{figure}

\begin{figure}
\epsscale{0.8}
\plotone{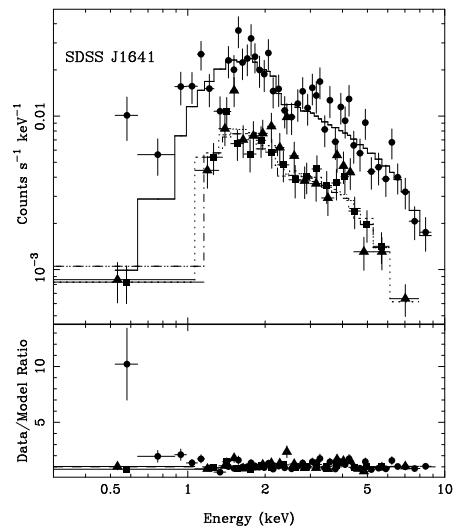}
\figcaption{\xmmn\ spectra of SDSS~J1641+3858. Marks and lines are the same as
  in Figure \ref{pic_0210}.  \label{pic_1641}} 
\end{figure}

The spectral fits are shown in Figures \ref{pic_0115} -
\ref{pic_1641}, and Table \ref{t:plcfits} displays the parameters derived
for the objects in our sample.  The fits are
statistically acceptable for all objects except SDSS~J1641+3858.
For five of the eight, we were able to constrain the column density of
absorption, finding $\sim 2 \times 10^{22}$~H~cm$^{-2}$ for four of them
and roughly ten times that amount in the fifth case (albeit with poor
statistics).  For four of
the eight we were able to constrain the continuum slope, and all four are
significantly harder than typical Type I AGN:
three have $\Gamma \simeq 1.4$, and the fourth has $\Gamma \simeq 0.5$.
After correction for light lost to absorption, the five spectral fits
indicate 2--10 keV luminosities between 0.8 and $7 \times 10^{44}$~erg~s$^{-1}$.

\begin{figure}
\epsscale{1.0}
\plottwo{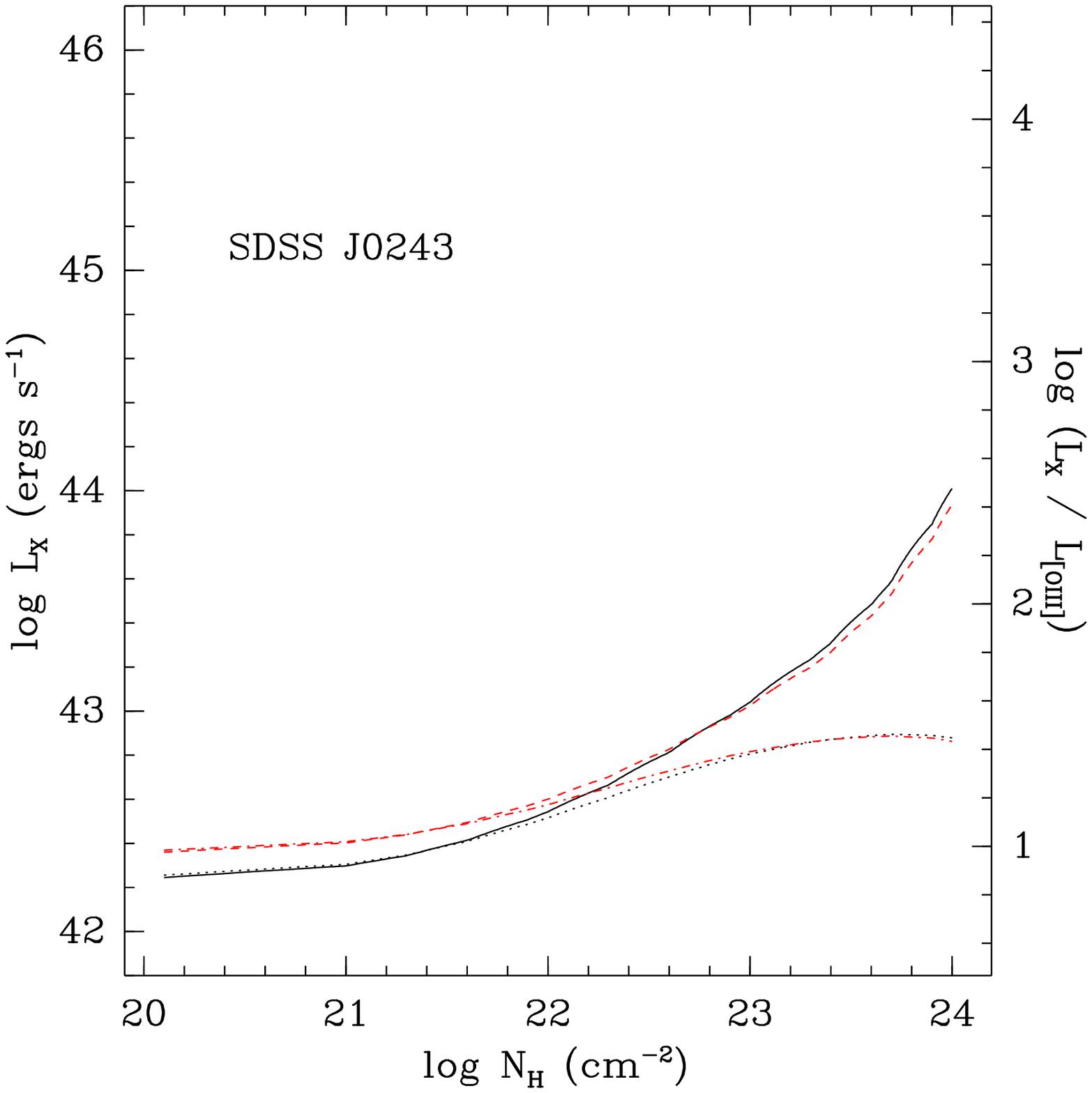}{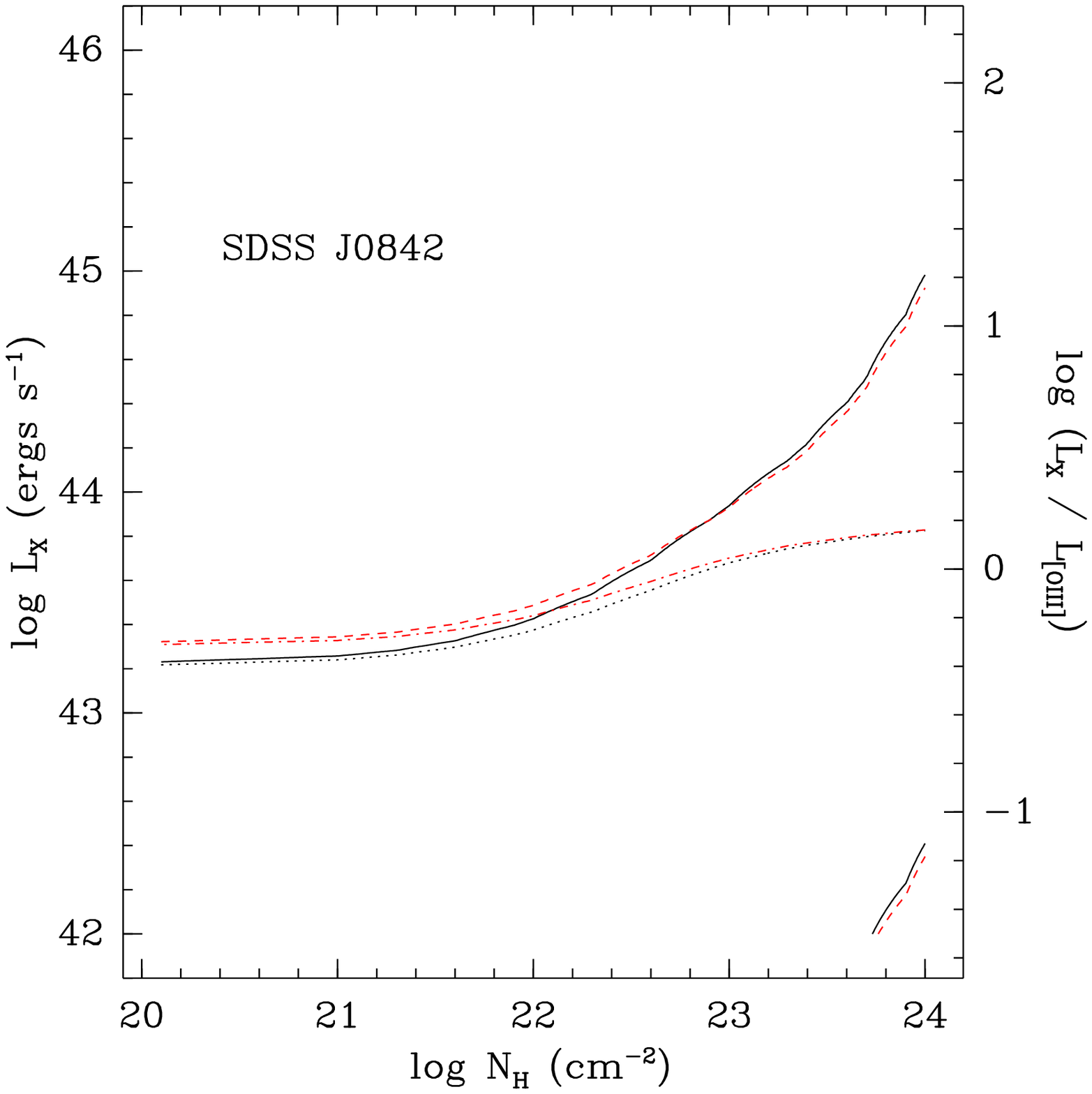}
\figcaption{The regions below and to the right of the curves are
  99.7\% confidence regions for X-ray luminosity (rest-frame $2-10$
  keV) and column
  density for SDSS~J0243+0006 (left) and SDSS~J0842+3625 (right).
  X-ray luminosity uncorrected for absorption is shown by
  dotted curves ($\Gamma=1.7$) and dot-dash curves ($\Gamma=1.4$).
  Absorption-corrected values are shown by solid curves ($\Gamma=1.7$)
  and dashed curves ($\Gamma=1.4$).  For additional contrast, the
  $\Gamma=1.4$ curves are colored red.
  Since these sources were not detected at high
  significance, these regions are generally unbounded
  from below (i.e., low luminosities are not excluded).
  In the case of SDSS~J0842+3625, extremely low luminosity and high
  column density is formally excluded because the probability
  that this object was actually detected (99.5\%) almost reached the 
  confidence level used in this figure (99.7\%).
  However, the range of column densities in question fall well
  into the regime where the approximations of the model (plcabs)
  are questionable.  We therefore treat these curves as showing
  only a luminosity upper limit.
  \label{f_upperlims}}
\end{figure}

For two objects, SDSS~J0243+0006 and SDSS~J0842+3625, we regard our
data as providing only upper limits on the X-ray flux because the
Poisson probability of a spurious source at those locations was
greater than 0.3\%.
Because SDSS~J0243+0006 was observed by XMM-Newton, each observation
resulted in three separate upper limits.  We therefore generalized the
approach of \citet{kraf91}, where upper limits are derived for
(single) observations in the Poisson statistical regime, to account for the
different responses of the MOS and PN detectors as described in
Appendix A. The results of this analysis are shown in Figure 
\ref{f_upperlims}, which shows 99.7\% confidence regions in the
$L_X-N_H$ plane for $\Gamma$ assumed to be either 1.4 or 1.7.  These 
regions are unbounded toward higher values of $N_H$ since no flux
would be expected if $N_H > 10^{24}$ cm$^{-2}$ (in the Compton-thick case, i.e., when the line-of-sight material is optically thick to Thompson 
scattering, \citealt{coma04}).
We plot the regions for both the absorbed and
intrinsic X-ray luminosities. For comparison with the spectral fitting 
results, in Table \ref{t:plcfits} we also list the upper limits 
corresponding to fixing the
column density at $10^{23} \rm \ cm^{-2}$.

We find it intriguing that the continuum slopes of the objects in our 
sample are consistently
shallower than in type I AGN, e.g., $\Gamma = 1.8 - 2.0$ in \
citet{risa05}, based on an XMM serendipitous survey of
photometrically-selected SDSS quasars.  It may be that this is
because the  
spectral complexity often seen in surveys of Seyfert 
II galaxies \citep{turn97, risa02} may be masked by our comparatively
poor photon statistics.  One possible complication is ``reprocessing",
in which X-ray photons are reflected from optically-thick material
located out of the line of sight (e.g., the accretion disk or the
opposing side of the putative torus) or scattered by optically-thin,
highly-ionized gas (e.g., the matter that creates the polarization
signature).   Alternatively, the obscuring 
material may not uniformly cover the X-ray source. In both cases, the
observed X-rays are absorbed by a range of column densities, which
tends to flatten the observed X-ray spectrum. 
 
To investigate this possibility, we have explored fitting a two-absorber
model (both at the redshift of the source) to the two sources with
the best photon statistics (SDSS~J1226+0131 and SDSS~J1641+3858).
That is, in the limit of negligible Compton scattering ($N_H \ll
10^{24} \rm \ cm^{-2}$), our model reduces to the form
\small
\begin{eqnarray*}
dN/dE = A \exp{\left[-\sigma(E) N_{\rm H,
      Gal}\right]}\left[E(1+z)\right]^{-\Gamma}  \times \\
   \left\{f\exp{\left[-\sigma(E\left[1+z\right]) N_{H,1}\right]} +
       \left(1-f\right)\exp{\left[-\sigma(E\left[1+z\right])
	   N_{H,2}\right]}\right\}
\end{eqnarray*}
\normalsize
where $E$ is the observed photon energy, $\sigma(E)$ is the
absorption cross section per H atom, $f$ 
gives the fraction of the source obscured by column density $N_{H,1}$,
while the remainder of the source is covered by matter with column
density $N_{H,2}$.
The two-absorber model is an approximation to either the case of
partial covering or the case of some X-ray flux being scattered into
the line of sight around a higher column-density absorber.
The results of the fits (i.e., the two column densities and the 
covering fraction
$f$ for the smaller column density) are shown in Table \ref{t:2plcfits}.
For both objects, the best-fit column density on the line-of-sight of
greater covering fraction increased, by a modest amount for SDSS~J1641+3858,
and by at least an order of magnitude for SDSS~J1226+0131.  In both
cases the smaller column density covers $\lesssim 10\%$ of the source.
As expected, the derived intrinsic continuum slope grew significantly steeper in both cases (for SDSS~J1641+3858, the resulting value of $\Gamma$ is nominally larger than in the single-absorber case, but is not well-constrained).
The results of fitting a two-absorber model should be treated with caution, 
because in the case of SDSS~J1226+0131
the single-absorber model is statistically satisfactory 
($\chi^2$/dof$=28.1/25$, which can be rejected at a confidence level
of only 70\%), so there is no requirement to introduce a more  
complex model in that case. However we note that these covering
fractions (1-f, or $>90\%$) are similar to the values observed in
nearby Seyfert 2 galaxies \citep{turn97}.  
On the other hand, in the case of SDSS~J1641+3858, 
while the quality of fit is improved, it is still not entirely satisfactory 
($\chi^2$/dof$=99.8/75$, which can be rejected at the 97\% confidence
level). We discuss 
this latter object in more detail in the following section.


\section{Notes on Individual Objects}\label{sec_individ}

{\bf SDSS~J0115+0015:}
This AGN is located behind the cluster Abell 168 and was covered
serendipitously by a \ch\ observation of this cluster, 1.75\arcmin\
off-axis. The X-ray source was detected with $\sim 300$
counts, with a negligible background ($\sim 2$ counts in the source
aperture of 2''). Both its [OIII]
luminosity ($\log (L$[OIII]$/L_{\odot})=8.14$) and the intrinsic X-ray
luminosity ($6\times 10^{43}$ erg s$^{-1}$) place this object
somewhat below the quasar luminosity cut-off (we adopted $\log
(L$[OIII]$/L_{\odot})>8.5$ as the optical definition of a quasar 
from Paper I, and $L_X>1\times 10^{44}$ erg s$^{-1}$ as the
X-ray definition, \citealt{szok04}). Nevertheless, this is still a
luminous AGN and is therefore included in the sample.  
In addition to the fit to the unbinned spectrum, we also
fit this spectrum after binning to 20 
counts bin$^{-1}$, which resulted in very similar fit parameters.  The
single-absorber fit (Figure \ref{pic_0115}) resulted in a column 
density of $2 \times 10^{22} \rm \ cm^{-2}$.  

{\bf SDSS~J0210-1001:}
This object was observed as part of our XMM-Newton program.  Its
observed column density is $2 \times 10^{22} \rm \ cm^{-2}$ (the absorbed 
power-law fit is shown in Figure \ref{pic_0210}), however
the observed spectral slope is very flat ($\Gamma = 0.5$).  This may
be a case in which reprocessing, patchy absorption or other effects are 
important (as described in Section \ref{sec_spec}), but there are insufficient
counts to justify any such model.

{\bf SDSS~J0243+0006:}
This object is within the field of view of two \xmmn\ observations of
NGC 1068, located $\sim 10$\arcmin\ off-axis.  Like SDSS~J0115+0015, its [OIII]
luminosity is not great enough for it to be labeled a quasar.  It
was not detected in any of the six exposures
(using a source aperture of 21\arcsec), and the upper limits on the
luminosity were obtained by combining the six exposures as discussed
above and in the Appendix (see Figure \ref{f_upperlims}).  Even summing
the counts, the nominal Poisson probability of no real source at this
location was 15\%.

{\bf SDSS~J0801+4412:}
This object was detected with $\sim 40$ counts in a pointed \ch\
observation from our program, too few to constrain the continuum
slope.  Less than $0.1$
background counts are expected in the 
source aperture, so we used the C statistic with no binning and
without background subtraction in our spectral fitting (Figure \ref{pic_0801}).
The resultant column density was
$\sim 1 \times 10^{23} \ \rm cm^{-2}$ for assumed photon indices of
either 1.4 or 1.7.

{\bf SDSS~J0842+3625:}
This object has the second highest [OIII] luminosity,
after IRAS~09104+4109, of all known type II quasars  \citep{klei88,
craw96, fran00, iwas01} at redshifts $z\la 1$.  Like IRAS~09104
\citep{hine93}, SDSS~J0842+3625 is 16\% polarized in the
optical \citep{zaka05}.  Unlike IRAS~09104, which is a double-lobed
radio source, SDSS~J0842+3625 has only a weak 
point-like radio counterpart and is radio-quiet, as determined by the
radio-to-[OIII] ratio. The source was imaged by \ch\ ACIS-I on two
occasions, in Oct~1999 (observation 
ID~532, 8 ks exposure) and in Dec~2002 (observation ID~4217, 20 ks exposure)
because it lies 6\arcmin\ away from the center
of the cluster Abell 697.   Unfortunately, in the
latter exposure SDSS~J0842+3625 happens to lie between the ACIS-I
detectors; we therefore discuss only the 8 ks observation.
In that image, SDSS~J0842+3625 is 6\arcmin\ off-axis, where the
point-spread function of \ch\ is rather larger than on-axis.
We used a 4\arcsec\ extraction radius around the SDSS position, in
which we found six counts.  To estimate the background, we used two
circular regions 25'' in radius offset from the source position, from
which we estimated there would be 1.0 $\pm$ 0.2 and 1.5 $\pm$ 0.2
background counts in the source region.  This variation, while significant at
less than the 2$\sigma$ level, may be due to the cluster emission.  The
count-to-background ratio from the higher background estimate implies that the
source is detected with 99.5\% confidence \citep{kraf91}.  We
therefore conservatively consider the observation to be an upper-limit
(see Figure \ref{f_upperlims}).  This source is described as
undetected by \citet{vign04} based on the same data. 

{\bf SDSS~J1226+0131:}
This object was serendipitously covered by an \xmmn\ observation of an
extragalactic gas-rich system, HI1225+01. A significant fraction of the
exposure ($\sim 50\%$) was affected by strong flaring, and the data
collected during the flares were excluded from the analysis. The single-absorber fit resulted in a column density of $N_H = 2\times 10^{22}$ cm$^{-2}$ (Figure \ref{pic_1226}). The spectrum
of this object was also presented by \citet{vign04}, who obtained
similar spectral parameters.

\begin{figure}
\epsscale{1.0}
\plotone{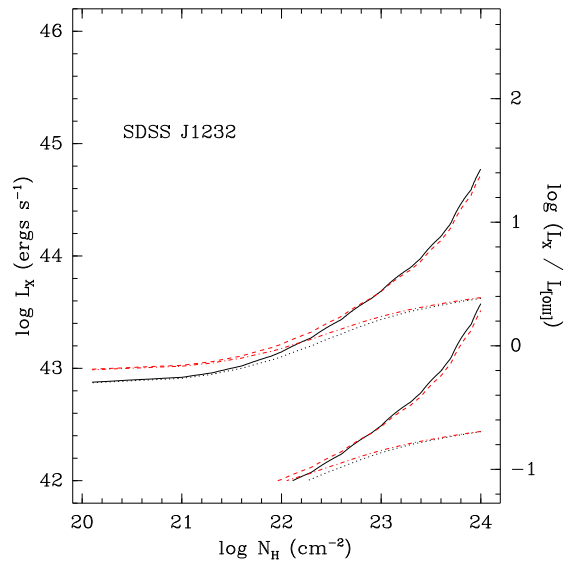}
\figcaption{99.7\% confidence regions for  X-ray luminosity (rest-frame $2-10$
  keV) and column density for SDSS~J1232+0206. 
  X-ray luminosity uncorrected for absorption is shown by
  dotted curves ($\Gamma=1.7$) and dot-dash curves ($\Gamma=1.4$).
  Absorption-corrected values are shown by solid curves ($\Gamma=1.7$)
  and dashed curves ($\Gamma=1.4$).  For additional contrast, the
  $\Gamma=1.4$ curves are colored red. As discussed in the text, this
  source was detected at $>99.7\%$ confidence and therefore the
  confidence region is bound from below (i.e., very low luminosities
  are excluded).
  \label{f_1232}
}
\end{figure}

{\bf SDSS~J1232+0206:}
This object was a target in our \ch\ program.  Six counts were detected
within a 2'' aperture at the position of the source and the background
estimate in the source aperture is $0.3$ counts.  This means
that the source was marginally detected (Poisson
probability of a false detection $1 \times 10^{-6}$).  Since we
have only barely enough counts for a detection, we follow the procedure
discussed in the Appendix.  We find that the 99.7\% confidence
contours (equivalent to $3\sigma$ in Gaussian statistics) define
a band in the luminosity-column density plane
(Figure \ref{f_1232}).  Because it is detected, albeit weakly,
we can exclude the very low luminosity and high column density
corner of this plane.  We do not consider
X-ray luminosities below $10^{42} \rm \ erg \ s^{-1}$ because
starburst X-ray emission can dominate over low-luminosity AGN
emission at this level (e.g., see Ptak et al. 1998).
In
addition, we demonstrated in our earlier work (Zakamska et al. 2005)
that scattering in the interstellar material in the host galaxy on kpc
scales can be a significant effect, and around 1\% of the intrinsic
optical emission can thus reach the observer. If scattering is
primarily by electrons, then the same scattering efficiency applies to
the X-ray emission, and an apparent X-ray luminosity of $10^{42}$ erg
s$^{-1}$ can be dominated by the scattered light that has not been
subject to circumnuclear obscuration.
Therefore, for $N_H \la 10^{22} \rm \ cm^{-2}$, our upper limit
effectively becomes $L_X$ $\la 10^{43}$~erg~s$^{-1}$. 
This object was listed by Zakamska et al. (2004) as having a potential
counterpart in the RASS catalog, with a nominal offset between the
SDSS position and the RASS position of 0.75\arcmin. Our {\it Chandra}
data indicate that this is likely a mismatch, as there is a much
brighter source in the field closer to the nominal RASS position.  

\begin{figure}
\plotone{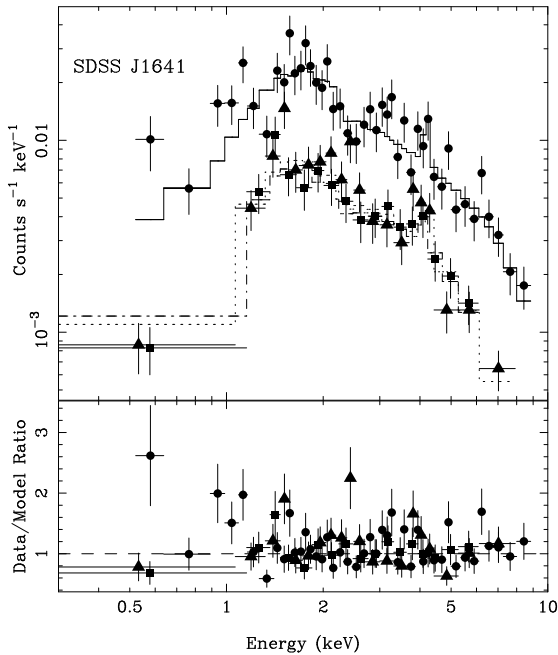}
\caption{The \xmmn\ spectra of SDSS~J1641+3858 fit with a power-law plus
  two absorbers and a narrow Gaussian.  The Gaussian line energy was
  consistent with a rest-frame energy of $\sim 6.7$ keV. Note that this fit is
  marginally statistically acceptable (rejected at a confidence level
  of 93\%) while the single-absorber
  power-law fit is clearly not (rejected at a confidence level of 99.5\%).\label{f:1641_2plcabs}}
\end{figure}

{\bf SDSS~J1641+3858:}
This object was observed by {\it XMM-Newton} as part of our program, and 
$\sim 1600$ counts were detected. As mentioned above, a single absorber
fit (Figure \ref{pic_1641}) is statistically unacceptable. The addition of a 
second absorber
resulted in a marginally acceptable fit. 
Residuals remained at an
observed energy of $\sim 4.0$ keV, which corresponds to 6.4 keV in the
rest frame, the energy of the neutral Fe K$\alpha$ emission line. We
added a narrow (0.01 keV physical width) Gaussian to the fit, 
which reduced $\chi^2$ by $\sim 8$ for 2 additional parameters, but this
does only slightly improves the confidence level of the fit. The
rest-frame energy of the line was 6.69 (6.60-6.80) keV and the
equivalent width (EW) was 0.16 (0.05-0.35) keV.  This line energy is
signicantly higher than 6.4 keV, which is expected from ``neutral'' Fe
(less ionized than Fe XVII) and is consistent with He-like
Fe-K$\alpha$ emission.  \citet{nand97} and \citet{reev00} found
evidence for a luminosity dependence in the Fe-K emission of AGN,
possibly due to the putative accretion disk becoming ionized at
high accretion rate.  However, \citet{jime05} found no correlation
between Fe-K line energy and X-ray luminosity in a sample of PG
quasars.  The observed EW is similar to values observed in Type-1
radio-quiet AGN \citep{reev00}, suggesting that the continuum we are
observing at 6 keV is not obscured.  This fit is shown in Figure
\ref{f:1641_2plcabs}.

\section{Discussion}
\label{sec_disc}

The primary goal of our study was to determine the X-ray spectral
characteristics of a sample of type II quasars which were selected 
based on their narrow emission lines.  Because our sources were all
identified exclusively on the basis of optical properties, our sample
should be free of any X-ray selection bias.  The four in our
program were chosen as those having the greatest [OIII] flux;
SDSS~J0842+3625 would have been in our program but for the fact
that it had already been observed in a cluster observation;
the other three were observed serendipitously.

\subsection{The column density distribution}

All five of our type II quasars with enough counts to constrain
absorption showed column densities of at least $\sim 10^{22}$~cm$^{-2}$.
As we will argue below on the basis of its X-ray/[OIII] luminosity ratio,
we believe it likely that SDSS~J0842+3625 was undetectable
because its obscuration is Compton-thick.  Thus, while our
sample is small, we have nonetheless shown
it is very likely that type II quasars are in general strongly
absorbed.


Our sample can be compared to X-ray observations of nearby Seyfert II
galaxies, with the caveat that archival X-ray surveys of nearby AGN
suffer from an X-ray selection bias because the AGN were usually
observed at least in part on the basis 
of their X-ray flux.  In addition, it should be kept in mind
that at lower luminosities the soft X-ray band can be dominated by
starburst emission.  Several such surveys exist in the literature.

In an {\it ASCA} survey of Seyfert II galaxies, \citet{turn97}
found somewhat steeper continua, but very similar column
densities, to those encountered in our sample: the mean power-law
index and absorbing column in their study were
$\Gamma=1.8$ and $N_H = 4 \times 10^{22} \rm \ cm^{-2}$. However, often the
absorbed power law fits were poor.  Fitting their data with more complex
models, such as partial covering with neutral or ionized
material, a thermal plasma component (representing starburst
emission below 2 keV), or reflection (off neutral, Compton-thick
material out of the line of sight), improved the fits.  In
the case of partial covering by neutral material, the mean photon
index increased to $\sim 2.0$ and the mean column density increased to
$\sim 1 \times 10^{23} \rm \ cm^{-2}$, although in a number of cases a
double power-law model (in which the two power-laws were allowed to
have different slopes and the absorption for one power-law was fixed at
the Galactic value) provided better fits.

\citet{risa02} presented a {\it BeppoSAX} survey of Compton-thin Seyfert II
galaxies, where the simplest model fit to the data was a double
power-law model.  The mean
spectral index and column density (for the absorbed power-law) from this
analysis were $\Gamma \sim 1.8$ and $N_H \sim 2 \times 10^{23} \rm \
cm^{-2}$, similar to the {\it ASCA} results.  Note that the
partial-covering fits to the ASCA data resulted in a slightly steeper
slope ($\Gamma \sim 2$) than found here, probably due to the inclusion
of PDS data at higher energies, where reflection effects become
significant. 

The previous work most nearly comparable to ours was that of
\citet{risa99}, who studied the X-ray column density distribution of
a sample of type II Seyfert galaxies, likewise selected on the basis
of [OIII] flux.   These authors found that 75\% had column
densities $N_H > 10^{23}$~cm$^{-2}$ and 50\% were Compton-thick.
Thus, if anything, the evidence in hand suggests that while the
incidence of absorption in type II Seyferts and quasars is similar,
its thickness may be somewhat greater in the Seyferts. However the
statistics are very limited and a larger sample is necessary to confirm
the trend and to determine whether this
contrast (if real) is primarily due to luminosity or redshift.


\subsection{Spectral slopes}

We were able to constrain the continuum slope in four objects and
achieve an acceptable fit for three of the spectra.  As we have already
remarked, in general we find rather harder continuum slopes than
are commonly found in type I AGN, whether Seyfert galaxies or
quasars.  As mentioned in
the previous subsection, our slopes are also harder than those
found in previous studies of type II Seyfert galaxies.

This contrast in slopes may not be genuine.  Studies of
type II Seyfert galaxies with better photon statistics have
shown that more complicated models than a power-law with
a single absorber are often required to adequately
describe the data \citep{turn97}.  

By fitting the two objects with the best statistics with more
complex spectral models, we have seen indications that these
hard spectra (and possibly also our modest column densities)
are indeed artifacts of fitting low signal-to-noise
data with a model simpler than is appropriate.  However, we
do not have sufficient signal to properly test this speculation.

\subsection{[OIII] emission as an indicator of X-ray luminosity}

\begin{figure}
\plotone{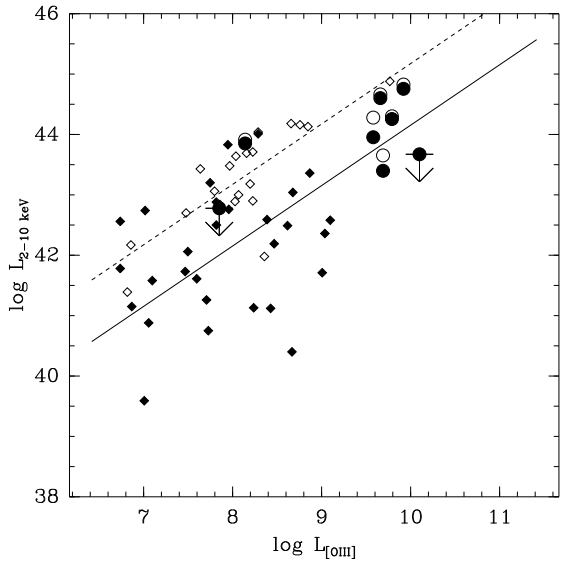}
\caption{2-10 keV X-ray luminosity (in units of ergs s$^{-1}$) plotted
 as a function of 
 $L$[OIII] (in units of $L_{\odot}$).
 Seyfert I (empty diamonds) and Seyfert II (filled diamonds) samples
 were taken from \citet{heck05}. The type II quasars from this paper
 are shown with filled and empty circles showing the X-ray luminosity
 before and after correcting for absorption (as listed in Table 2), i.e., in
 both the Seyfert 
 and our AGN samples, the empty symbols should represent intrinsic
 X-ray luminosities. The dashed and solid lines show the mean $\log
 (L_X/L$[OIII]) values from \citet{heck05} for the [OIII]-selected
 Seyfert I and II 
 samples (1.6 and 0.6 dex, respectively; neither $L_X$ nor $L$[OIII]
 were corrected for absorption in these samples). The upper-limits 
 are based on assumed column densities of $10^{23} \ \rm cm^{-2}$, and
 the corresponding limits on intrinsic luminosities are a factor of $\sim 2$
 higher than shown. \label{f:tim_oiii}
}
\end{figure}

Tables \ref{t:plcfits} and \ref{t:2plcfits} list the $L_X/L$[OIII]
ratios. The [OIII] luminosities of the objects in our sample are 
not corrected for possible reddening of the narrow-line region, since
the H$\alpha$ emission line is redshifted out of the optical range for
all objects in our sample, and thus we are not able to compute the
narrow-line Balmer decrement.  These ratios may be compared
with those found in the study of \citet{heck05}, who analyzed
the hard X-ray and [OIII] luminosities of nearby AGN.  They found that
when a sample of type II Seyferts is selected by [OIII] flux, the mean
$\log_{10}(L_X/L$[OIII]$) = 0.6$ with standard deviation 1.1.
All our detected objects (and the upper limit for one of our
undetected objects) fall well within the extrapolation of this
distribution to higher luminosities (Figure ~\ref{f:tim_oiii}).
In this respect, our type II quasars fall nicely in line with
the behavior of type II Seyfert galaxies.

\citet{heck05} found in addition that the mean
X-ray/[OIII] ratio of hard X-ray-selected type II Seyferts is rather
larger than when the selection is based on [OIII] flux, and is then quite
similar to the ratio for similarly-selected type I Seyferts:
$\langle \log(L_X/L$[OIII]$)\rangle = 2.2$.  The most natural
interpretation of this result is that many type II Seyferts identifiable
through [OIII] emission are {\it absent} from hard X-ray-selected
samples, presumably because even hard X-rays are entirely blocked
by Compton-thick obscuration. A similar conclusion was found by
\citet{bass99}, where it was shown that the Fe-K emission line
equivalent width  in Seyfert 2 galaxies increases strongly with
decreasing X-ray/[OIII] ratio, with Compton-thin AGN mostly exhibitng
X-ray/[OIII] $> 1$.  This also suggests that the X-ray
continuum is largely becoming absorbed when the observed X-ray/[OIII]
ratio is low.
Because our upper bound on $L_X/L$[OIII]
for SDSS~J0842+3625 (assuming a column density of $10^{23}$~cm$^{-2}$)
is nearly two orders of magnitude below the
typical ratio for hard X-ray selected AGN, we suggest that it may be
such a case of Compton-thick obscuration.

In fact, although our detected objects are roughly consistent with
the distribution of $L_X$/$L$[OIII] ratios for [OIII]-selected type II
AGN, they fall systematically below the mean relationship for
X-ray selected AGN.  This, too, suggests that we may be underestimating
the true absorption even in these cases.

\section{Summary \label{sum}}
Our results have in large part confirmed that the standard
unification model satisfactorily describes optically-selected type II
quasars. Six of the 
eight are strong X-ray sources, and all five with enough counts to fit
spectra show substantial absorbing columns.  Outstanding questions
include: Is the contrast in continuum slopes that we find between
typical type I quasars and these objects real?  Is the column density
distribution of type II AGN a function of redshift or luminosity,
as hinted by this small sample?  Observations of larger samples
with greater signal-to-noise ratios will be necessary to make progress.

\acknowledgements

We thank the referee, Dr. Roberto Della Ceca, for a very careful review.
Funding for the creation and distribution of the SDSS Archive has been provided by the Alfred P. Sloan Foundation, the Participating Institutions, the National Aeronautics and Space Administration, the National Science Foundation, the U.S. Department of Energy, the Japanese Monbukagakusho, and the Max Planck Society. The SDSS Web site is http://www.sdss.org/. 

The SDSS is managed by the Astrophysical Research Consortium (ARC) for the Participating Institutions. The Participating Institutions are The University of Chicago, Fermilab, the Institute for Advanced Study, the Japan Participation Group, the Johns Hopkins University, the Korean Scientist Group, the Los Alamos National Laboratory, the Max-Planck-Institute for Astronomy (MPIA), the Max-Planck-Institute for Astrophysics (MPA), New Mexico State University, University of Pittsburgh, University of Portsmouth, Princeton University, the United States Naval Observatory, and the University of Washington.

This research has made use of the High Energy Astrophysics Science
Archive Research Center (HEASARC, http://heasarc.gsfc.nasa.gov/)
operated by the Laboratory for High Energy Astrophysics at NASA/GSFC
and the High Energy Astrophysics Division of the Smithsonian
Astrophysical Observatory.  

AFP acknowledges the support of NASA grants NNG04GF79G and
GO4-5129X. NLZ and MAS acknowledge the support of NSF grant
AST-0307409. NLZ also acknowledges the support of the Charlotte
Elizabeth Procter Fellowship.

\begin{appendix}
\section{Multi-Detector Luminosity and Column Density Limits}

As discussed in Kraft et al. (1991) and Van Dyk et al. (2001),
Bayesian statistics are particularly useful for the interpretation of
data with low numbers of counts.  We begin by
computing the Bayesian posterior probability for X-ray luminosity
($L_X$), column density ($N_H$), photon index $\Gamma$ and background
within the source aperture ($B$), based on flat (i.e., constant)
priors on $L_X$ and $N_H$:  
\begin{equation}
P(L_X, N_H, \Gamma, B_{1..n} | D ) = \\
C P(\Gamma)\prod_{i}^{n} \mathcal{L}(T_i| S_i(L_X,
N_H, \Gamma) + B_i) \mathcal{L}(T_{B,i}| B_i A_b/A_s)
\end{equation}
where $\mathcal{L}(T_i| S_i(L_X, N_H, \Gamma) + B_{1..n})$ is the likelihood
  of observing $T_i$ total counts in the 
  source aperture from observation $i$, given a count estimate $S_i + B_i$, 
  and $\mathcal{L}(T_{B,i}| B_i A_b/A_s)$ is the likelihood of
  observing $T_{B,i}$ counts in the 
  background region (with area $A_b$) when the background estimate is
  $B$ in the source region (with area $A_s$).  The product is over the
  individual observations $i$, with each XMM detector being considered
  a separate observation since the responses differ.  Both likelihoods
  are Poisson distributions. 
 $S_i$ is the number of counts expected given the spectral model
 parameters, which is computed for the response (calculated at the
  position of the source) using {\sc xspec}.  $P(\Gamma)$ is the assumed
  prior distribution of photon indices, however here we fix $\Gamma$
  at the values of 1.7 and 1.4 (i.e., equivalent to a
  $\delta$-function prior).

We estimate $B_i$ 
using an annulus around the source or a source-free region close to
the source.
C normalizes the posterior probability to unity. The background counts estimates
$B_{1..n}$ are nuisance parameters which can be marginalized from the posterior
(hereafter dropping the $|D$ notation for the posterior):
\begin{equation}
P(L_X, N_H, \Gamma) \\
= \int_0^{\infty} dB_1 \int_0^{\infty} dB_2 \cdots
\int_0^{\infty}dB_n P(L_X, N_H, \Gamma, B_{1..n}) \\
= \prod_i^n
\int_0^{\infty} dB_i P(L_X, N_H,\Gamma, B_i)
\end{equation}   
The posterior can then be used to compute confidence contours in the
$L_X - N_H$ plane, which would be the smallest regions that encompass
the desired confidence level.
In practice, when we have an upper limit on 
the count rate, the most probable values will be near the highest
allowed $N_H$ and lowest allowed $L_X$, and the confidence contour
will be unbounded.  
\end{appendix}

\clearpage
\setcounter{table}{1}
\begin{landscape}
\begin{deluxetable}{cccccccc}
\tablewidth{0pt}
\tabletypesize{\scriptsize}
\tablecaption{SDSS type II AGN observed with \ch\ or \xmmn \label{t:params}}
\tablehead{J2000 & Galactic $N_H$ & & & Observation & Exposure &
  Date & off-axis \\ 
coordinates & ($\times 10^{20}$ cm$^{-2}$) & z & $\log
(L$[OIII]/$L_{\odot})$ & ID & (ks) & mm/dd/yy & angle (\arcmin)} 
\startdata
SDSS~J011522.19+001518.5 & 3.4 & 0.390 & 8.14 & \ch-3204 & 37.6 & 11/01/02 & 2.0\\
SDSS~J021047.01$-$100152.9 & 2.2 & 0.540 & 9.79 & \xmm-0204340201 & 9.7(P),11.6(M1),11.6(M2) & 01/12/04 & \\ 
SDSS~J024309.79+000640.3 & 3.6 & 0.414 & 7.95 & \xmm-0111200101 & 35.3(P),38.7(M1),35.6(M2) & 07/29/00 & 10.8\\
 &  &  &  & \xmm-0111200201 & 33.0(P),37.8(M1),34.9(M2) & 07/30/00 & 10.4\\
SDSS~J080154.24+441234.0 & 4.8 & 0.556 & 9.58 & \ch-5248 & 9.8 & 11/27/03 & \\
SDSS~J084234.94+362503.1 & 3.4 & 0.561 & 10.10 & \ch-532 & 7.3 & 10/21/99 & 5.4\\
SDSS~J122656.48+013124.3 & 1.8 & 0.732 & 9.66 & \xmm-0110990201 & 9.0(P),15.3(M1),15.4(M2) & 06/23/01 & 
6.0\\
SDSS~J123215.81+020610.0 & 1.8 & 0.480 & 9.69 & \ch-4911 & 9.5 & 04/20/05 & \\
SDSS~J164131.73+385840.9 & 1.2 & 0.596 & 9.92 & \xmm-0204340101 & 12.2(P),16.8(M1),17.1(M2) & 08/20/04 & \\
\enddata

\tablecomments{J2000 coordinates, redshifts and [OIII]$\lambda$5007\AA\
  luminosities are from Paper I. The Galactic absorption was derived
  from the HI map by \citet{dick90} using the {\sc nh} tool provided by
  the HEASARC. For \xmmn\ observations, the exposure times are listed
  separately for PN (P) and MOS1,2 (M1,2) instruments. The date of the
  observation is given in the next column. The off-axis angle is given
  in arcmin only for those four objects that were observed
  serendipitously. The remaining four objects were the primary targets
  of the observations.}
\end{deluxetable}
\clearpage
\end{landscape}
\clearpage
\begin{landscape}
\begin{deluxetable}{ccccccccc}
\tablewidth{0pt}
\tabletypesize{\scriptsize}
\tablecaption{X-ray spectral properties of SDSS type II AGN \label{t:plcfits}}
\tablehead{ID & Counts & $N_H$ & $\Gamma$
  & $\chi^2$/dof & $L_X$  &
  $L_{X,intr}$  & $L_X/L$[OIII] &
  $L_{X,intr}/L$[OIII]\\
& &  ($10^{22}$ cm$^{-2}$) & & & ($10^{44} \ \rm erg\ s^{-1}$) & ($10^{44} \ \rm erg\ s^{-1}$)}
\startdata
SDSS~J0115+0015 & 339 & 2.0 (1.3-2.8) & 1.39 (1.09-1.73) & 471/523 & 0.71 & 0.82 &  134 & 155\\
SDSS~J0210$-$1001 & 300 & 2.3 (0.4-5.5) & 0.46 (-0.07 -- 0.72) & 9.1/9 &1.8 & 2.0 & 7.5 & 8.3 \\
SDSS~J0243+0006 &  & 10 (fixed) & 1.7 (fixed) & & $<0.060$ & $<0.10$ &
$<18$ & $<30$ \\
SDSS~J0801+4412 & 40 & 16 (9.7-27) & 1.7 (fixed) & 221.9/524\tablenotemark{*} & 0.87 & 1.9 & 6.0 & 13. \\
%
SDSS~J0842+3625 & $<15$ & 10 (fixed) & 1.7 (fixed) &  & $<0.47$ & $<0.81$ & $<1.0$ & $<1.7$ \\
SDSS~J1226+0131 & 574 & 2.0 (1.4-2.7) & 1.41 (1.16-1.69) & 28.1/25 & 4.0 & 4.6 & 23 & 26\\
SDSS~J1232+0206 & 5.7 & 10 (fixed) & 1.7 (fixed) & & 0.25 & 0.45 & 1.3 & 2.4 \\
SDSS~J1641+3858 & 1624 & 2.8 (2.2-3.4) & 1.33 (1.19-1.51) & 113.5/77 & 5.7 & 6.7& 18. & 21\\
\enddata

\tablecomments{
The counts given in the case of XMM data are the net
  counts from the three detectors combined. In the case of
  SDSS~J0243+0006, no counts value is given since the upper limit on counts
  combined from disparate detectors is not meaningful. The errors on fit
  parameters $N_H$ (hydrogen column 
  density at the redshift of the source) and $\Gamma$ (photon index,
  $dN/dE\propto E^{-\Gamma}$) are based on $\Delta \chi^2=2.7$. $L_X$
  is the rest-frame $2-10$ keV luminosity and $L_{X,intr}$ is the
  rest-frame $2-10$ keV luminosity after correction for
  absorption.}
\tablenotetext{*}{Fitting was performed using the C statistic with
  unbinned data.}
\end{deluxetable}
\clearpage
\end{landscape}

\clearpage
\begin{landscape}
\begin{deluxetable}{cccccccc}
\tablewidth{0pt}
\tabletypesize{\footnotesize}
\tablecaption{Partial-Covering Fits \label{t:2plcfits}}
\tablehead{ID & $N_{H,1}$ & $N_{H,2}$ & $\Gamma$ &
  $f$ & 
$\chi^2$/dof & $L_X$
 /$L_{X,intr}$  & $L_X/L$[OIII]/$L_{X,intr}/L$[OIII] \\
& ($10^{22}$ cm$^{-2}$)  & ($10^{22}$ cm$^{-2}$) & & & ($10^{44} \ \rm erg\ s^{-1}$)} 
\startdata
%
SDSS~J1226+0131 & 3.6 (3.1-5.1) & 87 (53-150) & 2.83 (1.96-3.81) &
0.091 & 19.6/23 & 4.0 / 40. & 23 / 230 \\
SDSS~J1641+3858 & 0.41 (0.10-1.8) & 4.6 (3.4-8.7) & 1.56 (1.39-1.78) &
0.071 & 99.8/75 & 5.7 / 7.6 & 17.9 / 23.8 \\
\enddata
\tablecomments{Results of fitting a model consisting of two absorbers
  plus a power-law to the highest signal-to-noise ratio spectra.
The quantity $f$ is the fraction of flux intercepted by the absorber
with the lower column density.  See text for details.}
\end{deluxetable}
\clearpage
\end{landscape}

\end{document}